\documentclass[reprint,
 amsmath,amssymb,
 aps,prl
]{revtex4-1}
\usepackage{graphicx}% Include figure files
\usepackage{dcolumn}% Align table columns on decimal point
\usepackage{bm}% bold math
\usepackage{xcolor}

\begin{document}

\preprint{APS/123-QED}

\title{Surface Furrowing Instability in Everting Soft Solids}% Force line breaks with \\

\author{Jonghyun Hwang}
\author{Mariana Altomare}
\author{Howard A. Stone}
 \email{Corresponding author.\\hastone@princeton.edu}
\affiliation{%
 Department of Mechanical and Aerospace Engineering, Princeton University, NJ 08544, USA
}%

\date{\today}% It is always \today, today,

\begin{abstract}
We report a surface instability observed during the extrusion of extremely soft elastic solids in confined geometries. Due to their unique rheological properties, these soft solids can migrate through narrow gaps by continuously everting the bulk material. The extrusion front spontaneously buckles in the direction transverse to the flow, resulting in a furrow-like morphology that deepens over time. We characterize the distinct features of this instability using experiments and theory and contrast the results with known elastic surface instabilities. Our study may provide insights into various processes involving the extrusion-like deformations of soft solids, including biomaterials.
\end{abstract}

\maketitle

Soft materials exhibit various morphological features by virtue of their ready deformability. Often, mechanical interactions in soft materials cause surface buckling instabilities, which manifest periodic wrinkles, localized creases, or folds~\cite{dervaux2012ARCMP, biot1963ApplSci}. These instabilities can be of significant interest for applications in engineering, such as controlled optical properties~\cite{kim2011NatMat,kim2010NatMat} or predicting buckling modes of thin films~\cite{bowden1998nature}, and also pose challenging theoretical questions.

Surface instabilities have been studied primarily in the context of compression-driven instabilities, which arise from processes such as biological growth~\cite{richman1975Science,li2011JMPS}, swelling~\cite{dervaux2011PRL,cai2010SoftMat}, or friction~\cite{grillet2002JRheol,schallamach1971wear}. Studies on elastic surface instabilities typically have involved soft solids whose shear moduli were $O(10^3)$ Pa or larger.

Recently, the mechanics of solids in an extremely soft limit, whose equilibrium storage moduli $G_0 = O(0.1 - 10)$ Pa, have gained attention, owing to their natural connection to soft, solid-like systems found in biology~\cite{vincent2015PRL,deng2006NatMat,gardel2004Science}. Because of the loosely connected network structure, these solids exhibit rate-, time-, and strain-dependent elasticity; similar rheological characteristics are found in imperfectly networked elastomers~\cite{curro1983Macromol,chaplain1968Nature}. Consequently, understanding the mechanical traits of these synthetic and biological ultrasoft solids is essential. 

This Letter presents a surface instability observed during the movement of ultrasoft solids within confinement; we are not aware of previous reports of the phenomenon documented here. We demonstrate that these solids can translocate through narrow gaps by continuously inverting their own body inside-out. This deformation process leads to the emergence of furrow-like surface grooves, forming primarily due to the solid gel's `inside-out' deformation, and distinct from previously known types of surface instabilities. We confirm this distinction theoretically and experimentally and offer a scaling analysis to explain the cause of the ‘furrowing’ instability.

\begin{figure*}
\includegraphics{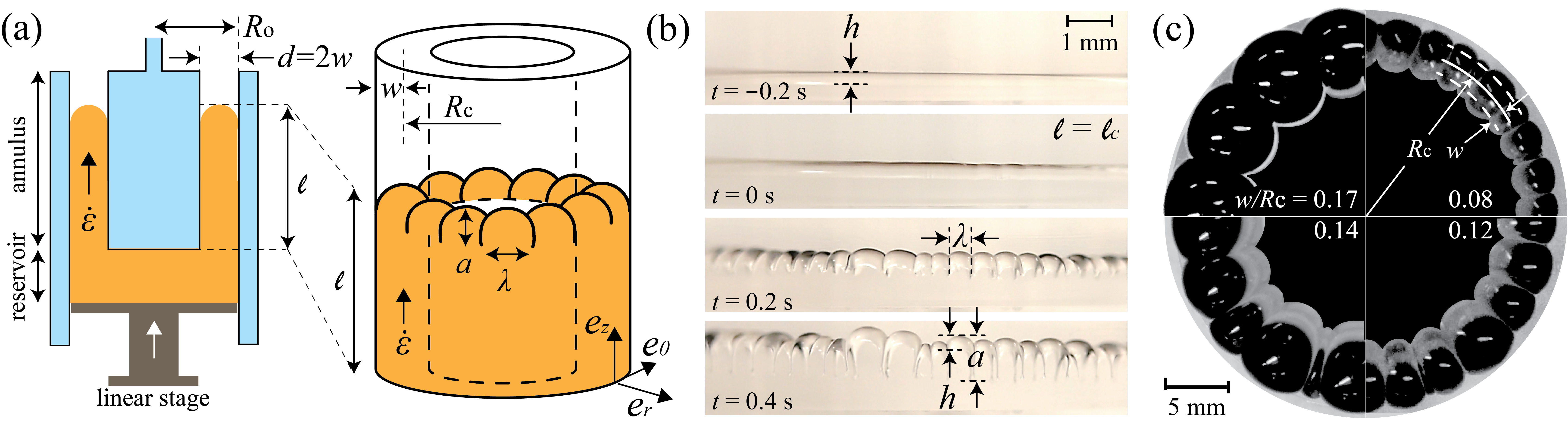}% Here is how to import EPS art
\caption{\label{fig:Figure1}Extrusion-driven surface instability. (a) Schematic of the experimental setup. The gel flows at strain rate $\dot\varepsilon$ in an annulus of a centerline radius $R_c$ and the gap $d=2w$. The location of the flow front $\ell$ is measured from the beginning of the annular section. The wavelength $\lambda$ along $R_c$ and furrow depth $a$ are measured at increasing $\ell$. (b) The initially flat gel-air interface becomes unstable at a critical length, $\ell_c$. The depth of furrows $a$ was measured from the apex. The height of the curved free surface, measured from the gel's contact with the wall, is denoted $h$. See Supporting video 1. (c) A top view of the fully extruded gel for different values of $w/R_c$. }
\end{figure*}

In our experiments, two acrylic cylinders, one hollow and the other solid, machined and surface-polished, were brought together to create a hollow annular channel of different widths $d=2w$; the centerline radius of the hollow annular channel is denoted $R_c$ [Fig.~\ref{fig:Figure1}(a)]. The outer radius of the annulus was $R_o=R_c+w=[7.9, 12.6, 15.5,21.9, 48]\pm0.3$ mm, and we adjusted the radius of the inner cylinder to give eleven different values of $w=[0.40,0.65,0.90,1.15,1.40,1.65,1.90,3.10,3.60,4.55,14.5]\pm0.08$ mm. Axisymmetric geometries were explored primarily due to having continuous boundaries to consider, whereas channels with rectangular cross-section introduce additional boundaries and corner effects; future experiments with long and narrow rectangular geometries may provide new insights. The viscoelastic gel material was cast and left to crosslink in the ‘reservoir’ below the annulus [Fig.~\ref{fig:Figure1}(a)].

The viscoelastic gel was composed of vinyl-terminated polydimethylsiloxane (PDMS) (viscosity $\mu=4850$ mPa·s; Gelest) and silicone oil ($\mu=20.8$ mPa·s; Sigma Aldrich). The vinyl group on PDMS reacts with a trimethylsiloxane terminated (15-18\% methylhydrosiloxane) dimethylsiloxane copolymer (Gelest) to form a crosslinked gel, and platinum-divinyl tetramethyldisiloxane complex 2\% Pt in xylene (Gelest) was used as the catalyst~\cite{hwang2023langmuir}. All materials were used as purchased. We explored three different crosslinking concentrations,  $c=[0.425,0.45,0.5]$ wt\% of the total weight of the PDMS and silicone oil mixture. 

The network components were mixed at non-stoichiometric ratios, thus creating dangling chains~\cite{mazurek2019ChemSocRev}. Dangling chains produce time-dependent (viscoelastic) material response in crosslinked gels. On short time scales, dangling chains entangle with the existing network, providing higher elastic resistance. On longer time scales, entangled dangling chains relax while the chemically crosslinked network provides the baseline elastic resistance~\cite{curro1983Macromol, ChassetThirion} (see~\cite{SupportingInformation}, \S~I). The storage modulus $G'$ in the long-time limit, which we denote $G_0$, was $O(0.1-10)$ Pa. 

Once the gel was fully cured, the gel was pushed vertically upward at a constant speed to flow (deform) into the hollow annulus (see~\cite{SupportingInformation}, \S~VIII). The position of the gel front $\ell$ was measured relative to the beginning of the annulus. The ratio $\ell/w$ provides a measure of axial strain and the average strain rate can be defined as $\dot\varepsilon = \ell/(wt)$, where $t$ is time, with $t=0$ when the gel first enters the annulus. $\dot\varepsilon$ was set by controlling the pump (PHD-Ultra, Harvard Apparatus), which was used as a linear translation stage. Once the gel front reached a certain length from the beginning of the annular section, the free surface of the gel spontaneously deformed, showing a periodic morphology [Fig.~\ref{fig:Figure1}(b)] (see~\cite{SupportingInformation}, \S~II). We denote this critical length as $\ell_c$ for the first appearance of this instability. The instability displayed larger morphological features with increasing $w/R_c$, as observed with top views of the gel [Fig.~\ref{fig:Figure1}(c)]. During the entire length of extrusion, no sign of gel fracture was observed. 

\begin{figure}
\includegraphics{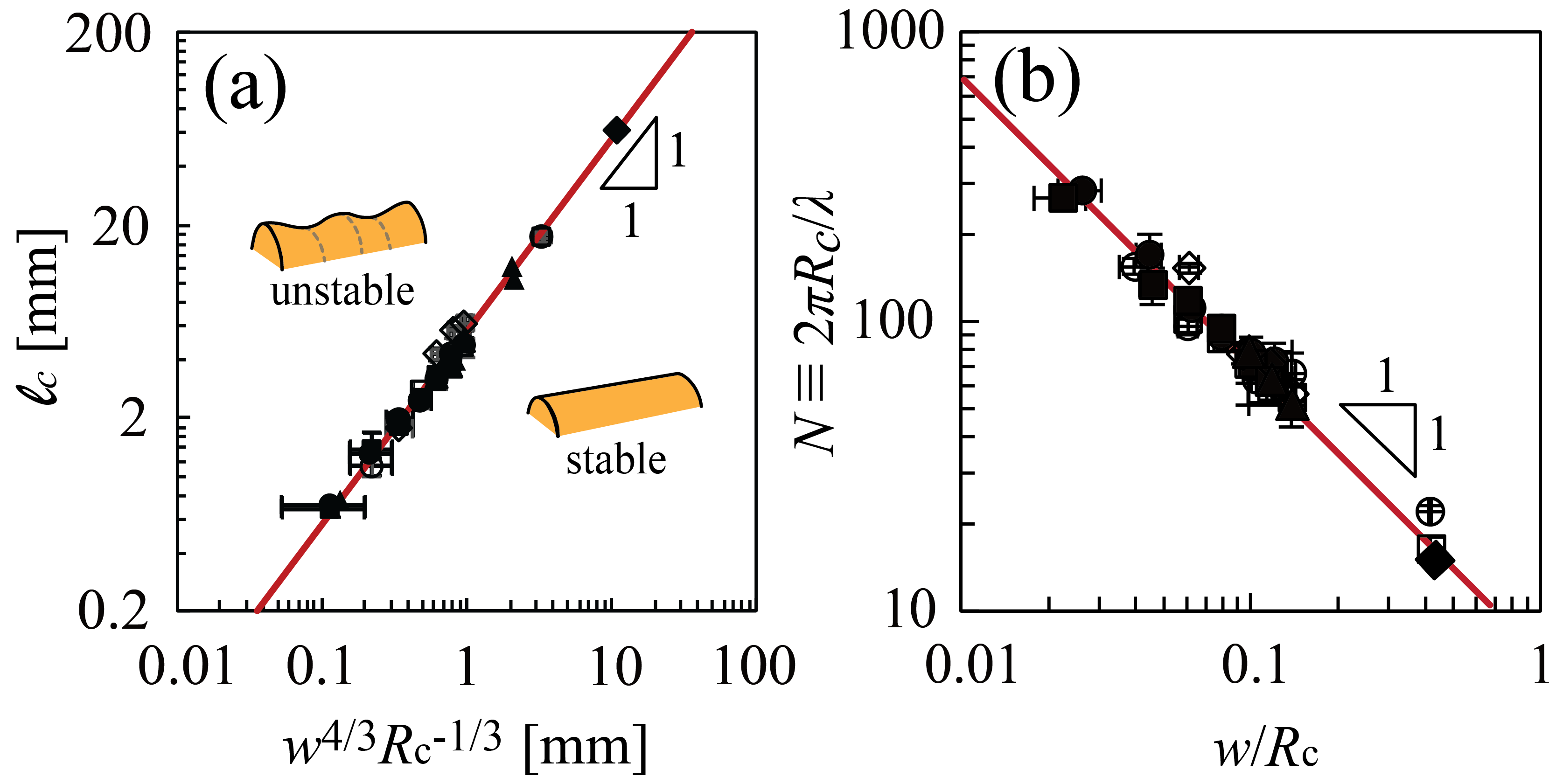}% Here is how to import EPS art
\caption{\label{fig:Figure2}Onset characteristics of the instability. (a) Critical length $\ell_c$ in relation to the geometrical length scales $w$ and $R_c$. Measurements show a power-law relationship $\ell_c\approx 5.4 w^{4/3}R_c^{-1/3}$ \cite{SupportingInformation}. (b) The number of fully formed protuberances $N$ right after the instability sets in shows $N\approx6.8 (w/R_c)^{-1}$, providing $\lambda\approx 0.93w$. Different symbols represent experiments performed with different combinations of $c$ and $\dot\varepsilon$. }
\end{figure}

We first discuss the onset characteristics of the instability. The critical length $\ell_c$ was measured at varying crosslinker concentration $c$, strain rate $\dot\varepsilon$, the half-gap $w$, and the outer radius $R_o$ [Fig. \ref{fig:Figure2}]. $\ell_c$ did not vary between different combinations of $c$ and $\dot\varepsilon$, however, it was most strongly affected by the geometrical length scale $w$. Within the range of $w$ studied, $\ell_c\propto w^{4/3}$. We explored the role of the confinement geometry more extensively and identified that $\ell_c\propto R_c^{-1/3}$; thus we could organize all the data as $\ell_c\sim w^{4/3}R_c^{-1/3}$ [Fig. \ref{fig:Figure2}(a)] (see \cite{SupportingInformation}, \S~III).

While the surface morphology of the instability shown in Fig.~\ref{fig:Figure1} resembles those observed in the compression-driven crease instability, we elucidate several key differences that set the instability documented here apart from the crease instability. The surface crease instability appears when a soft solid is compressed laterally greater than $\varepsilon_{crease}=35.4$ \% of its initial length~\cite{hohlfeld2011PRL}, or more if the effect of surface tension, $\gamma$, is significant. Localized creases are naturally aperiodic, but when the two length scales, the gel thickness, $w$, and the elastocapillary length, $\gamma/G_0$, are comparable, periodically buckled surface pattern can form~\cite{liu2019PRL}. In such cases, the wavelength and the critical compressive strain depends on both $w$ and $\gamma/G_0$~\cite{mora2011SoftMat}. Considering that the scale of the elastocapillary number that we investigated ranged $\gamma/wG_0=O(10^{-2}-10^2)$, critical compressive strains, if we are observing a crease-like instability, would need to be at least 35~\% or be larger than 75~\%~\cite{liu2019PRL}; however, compressive strains on the gel free surface in our annular geometry are less than 30 \% and typically smaller, as obtained by comparing the ratio of circumferences along $R_c$ and $R_i=R_c-w$.

Our experimental observations of the instability indicate distinct features from the crease instability. First, we showed that the instability still appeared without the gel reservoir, or even without the annular confinement, such that there is no overall compression of the elastic gel in the radial and circumferential directions (see~\cite{SupportingInformation}, \S~III). To show that, we experimented with hollow cylinders, not annulus, of varying outer radii $R_o=R_c+w=[3.8,5.2,8.0,10.8,15.5]\pm0.3$ mm. For hollow cylinders, we treated $R_i=R_c-w\rightarrow0$, and therefore, we examined if $\ell_c\sim w^{4/3}R_c^{-1/3}\sim w$. As expected, measurements showed $\ell_c\sim w$, confirming that compression is not the driving mechanism for this instability. Furthermore, our experiments show that both the critical transverse strain $\ell_c/w\approx(w/R_c)^{1/3}$ and the instability wavelength $\lambda=2\pi R_c/N\approx w$ depend only on the geometrical length scales $w$ and $R_c$, rather than being affected by $\gamma/G_0$ [Fig.~\ref{fig:Figure2}]. 

\begin{figure*}
\includegraphics[width=1\linewidth]{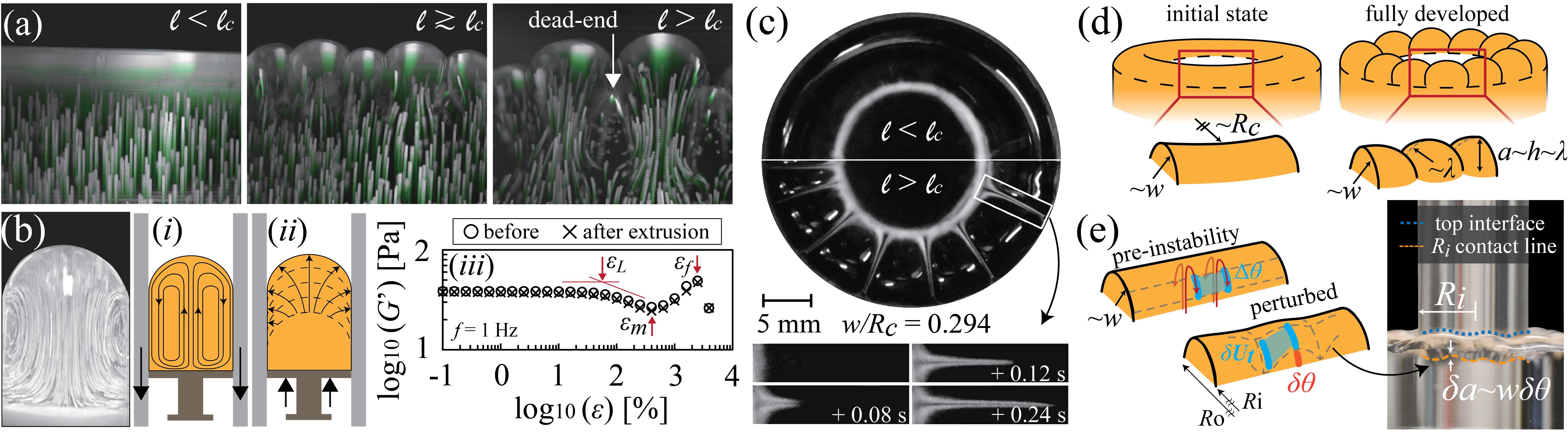}% Here is how to import EPS art
\caption{\label{fig: Figure3}Flow of soft solids in confinement. (a) Time-stacks of particle position; see Supporting video 2. (b) Depth-integrated image of particle trajectories inside everting soft solids. (\textit{i}) Streamline in the co-moving frame of reference and (\textit{ii}) laboratory frame of reference. (\textit{iii}) Amplitude-sweep measurement of the $c=0.45$ wt\% gel before and after the extrusion indicates that the network integrity was qualitatively preserved. (c) The fold line propagated from the inner to the outer boundary within 0.24 seconds for $c=0.5$ wt\% gel extruded at $\dot\varepsilon=0.1\text{ s}^{-1}$. The time scale of the instability growth is much faster than the time scale of the extrusion ($\dot\varepsilon^{-1}=10$ s, in this case) (see Supporting video 3). (d) The relevant length scales in the stable and unstable configurations. The fully developed state is defined as $a\sim h\sim\lambda$. (e) Perturbation in eversion angle $\delta\theta$ causes instability onset.}
\end{figure*}
We suspected that material confinement, i.e., interactions with the side walls, was a potential cause of the instability. Hence, we performed experiments after lubricating the side walls with thin layer of silicone oil ($\mu=5 \text{ mPa}\cdot\text{s}$). Such manipulation allowed the gel entering from the reservoir to slip on the wall of the annulus. Nevertheless, the gel experiences radial and circumferential compression as it deforms into a smaller area. The critical length for the instability increased to the prediction of the crease instability, which we confirm by performing a linear perturbation analysis (see~\cite{SupportingInformation}, \S~IV). 

To explore the deformation of the solid gel within confinement, we mixed density-matched polystyrene beads ($\rho=0.96$ g/cm$^3$, 106-125 µm-diameter; Cospheric), and then tracked the position of the particles [Fig.~\ref{fig: Figure3}(a)]. Once the surface was deformed, beads positioned underneath the furrowing folds stalled. As the gel progressed further into the annulus, some furrows grew, increasing $\lambda$. During this process, beads approaching the dead-end turned away and flowed into surviving, larger protuberances. 

The material trajectory of the elastic elements inside the curved tip region is similar to that of the fountain flow in viscous liquids, characterized by rotating toroidal vortices [Fig.~\ref{fig: Figure3}(b)]. We call this type of solid deformation `eversion'. This observation evinces that soft solids can move in confinement by constantly turning their body inside-out. We note that the instability begins and propagates from the inner boundary, not at the apex of the extrusion front [Fig.~\ref{fig: Figure3}(c)]. Also, the propagation of the instability happens much faster than the time scale of the extrusion ($\dot\varepsilon^{-1}$). This finite-time propagation is attributable to the gel viscoelasticity~\cite{hohlfeld2011PRL}. 

The everting path lines of the elastic gel element suggest that the instability observed here is shear-bending-driven, rather than compression-driven. This leads to the conclusion that the instability differs from the crease instability and represents a phenomenon that has not been explored before, at least to our knowledge. Henceforth, we refer to the instability studied in this work as the `furrowing' instability, drawing inspiration from its morphological resemblance to surface undulations that progressively deepen over time. Having this knowledge, we relate the relevant variables that characterize the eversion of soft solids to explain the observed power-law relationships; $\ell_c \sim w^{4/3}R_c^{-1/3}$ and $\lambda\sim w$. 

First, we experimentally observed that the height of the curved front $h$ linearly scales with $w$; $h\approx 0.8w$, independent of $\dot\varepsilon$ and $c$ (see Fig.~\ref{fig:Figure1}(b) to find $h$). This relationship is rationalized by modeling the progressing gel ‘column’ as a one-dimensional structure where the addition of the new material is localized in a top region with a constant height $h$ during extrusion~\cite{goriely2017Book} (see~\cite{SupportingInformation}, \S~V). Therefore, we consider the curvature of the gel front to scale with $ w^{-1}$.

The advancing gel front features two rotating (elastic) vortices within the curved tip, with material path lines following poloidal trajectories around the meridian. Because of the elastic nature of the material, we conceptualize the rotating vortices as two bending thin ($\sim w$) and wide ($\sim R_c$) sheets. To arrive at the tip, elastic materials experience shear interaction between the upward-flowing (away from the wall) and downward-flowing (near the wall) regions, accumulating the shear-induced strain energy $U_s$. Estimating the scale of strain as $\varepsilon\sim\ell/w$ and the sheared volume as $V\sim\ell w R_c$,  we obtain $U_s\sim G_0V\varepsilon^2\sim G_0\ell^3w^{-1}R_c$.

The confinement length scale governs the scale of bending energy. Indicating the length of the meridian $s\sim w\theta$, and the bending angle $\theta= O(1)$, we write the bending energy as  $U_b\sim BA\kappa^2$, where the bending stiffness $B \sim G_0w^3$, area $A\sim sR_c$, and curvature $\kappa=\partial\theta/\partial s\sim\theta/s$, from which we obtain $U_b\sim G_0 w^2 R_c$.  The scale of the bending energy does not increase with $\ell$ as $s$ is constant during the progression of the tip.

In the case of eversion inside an annulus, we consider the interplay of two principal curvatures ($\kappa_1\sim w^{-1}$ and $\kappa_2\sim R_c^{-1}$) introduced by the poloidal-toroidal geometry of the front. The material element moving along the poloidal trajectory experiences a sign change in the Gaussian curvature, which scales as $\kappa_G=\kappa_1\kappa_2\sim(wR_c)^{-1}$. Therefore, we write a Gaussian bending energy $U_{b,G}\sim BA\kappa_G\sim G_0w^3$~\cite{liang2009PNAS}, and thus the total energy is $U_b+U_{b,G}$. In contrast, the scale of the bending energy in the fully developed instability state, as sketched in Fig.~\ref{fig: Figure3}(d), with the furrow depth $a\sim \lambda$, can be written accounting for both Gaussian and mean curvatures ($\kappa'^2\sim w^{-2}+\lambda^{-2}+(w\lambda)^{-1}$) as  $NBA'\kappa'^2\sim G_0 w^2R_c \sim U_b$, where $A'\sim s\lambda$, $N\sim R_c/\lambda$, and $\lambda\sim w$ (experimentally observed).  Hence, we conclude that the furrowing instability lowers the total energy by $\sim U_{b,G}$ by breaking up a long ($\sim R_c$), rotating curved surface into a smaller ($\sim\lambda$) protuberances.

We then identify the critical extrusion length $\ell_c$ from where $U_s$ and $U_{b,G}$ become comparable; in other words,
\begin{equation}
G_0\ell^3w^{-1}R_c\sim G_0 w^3,
\end{equation}
arriving at $\ell\equiv\ell_c\sim w^{4/3}R_c^{-1/3}$. For $\ell<\ell_c$, the energetic cost associated with azimuthal bending dominates shear energy. Hence, surface elements are stable against perturbations, as plane shear extrusion is energetically favorable against bending in the azimuthal direction. If instead $\ell>\ell_c$, continued extrusion costs more energy as $U_s\propto\ell^3$, and, therefore, azimuthal bending due to perturbations becomes likely as it constitutes a reduced total energy. 

 As the gel progresses by rotating the material element along the free surface, the position of the gel element in the co-moving frame of reference [Fig. 3(b)] can be characterized by angular position, $\theta$. Therefore, emergence of the instability involves a perturbation to this degree of rotation, $\delta\theta$ [Fig. 3(e)]. This angular perturbation $\delta\theta$ produces torsional energy $\delta U_t$ within the gel front, $\delta U_t\sim G_0J\delta\theta^2/L\sim G_0 w^2\delta a^2/\lambda$, where the torsional moment of inertia scales $J\sim w^4$, the length scale of torsion scales $L\sim\lambda$, emerging furrow depth at the gel-wall contact line scales as $\delta a\sim w\delta\theta$. Likewise, $\delta a$ causes azimuthal bending $\delta U_b\sim G_0 w^5\delta a^2/\lambda^4$. As the two energy scales must balance in order to minimize the perturbed state of the energy, we balance
\begin{equation}
G_0 w^2\delta a^2/\lambda\sim G_0 w^5\delta a^2/\lambda^4,
\end{equation}
and arrive at $\lambda\sim w$.  

\begin{figure}
\includegraphics{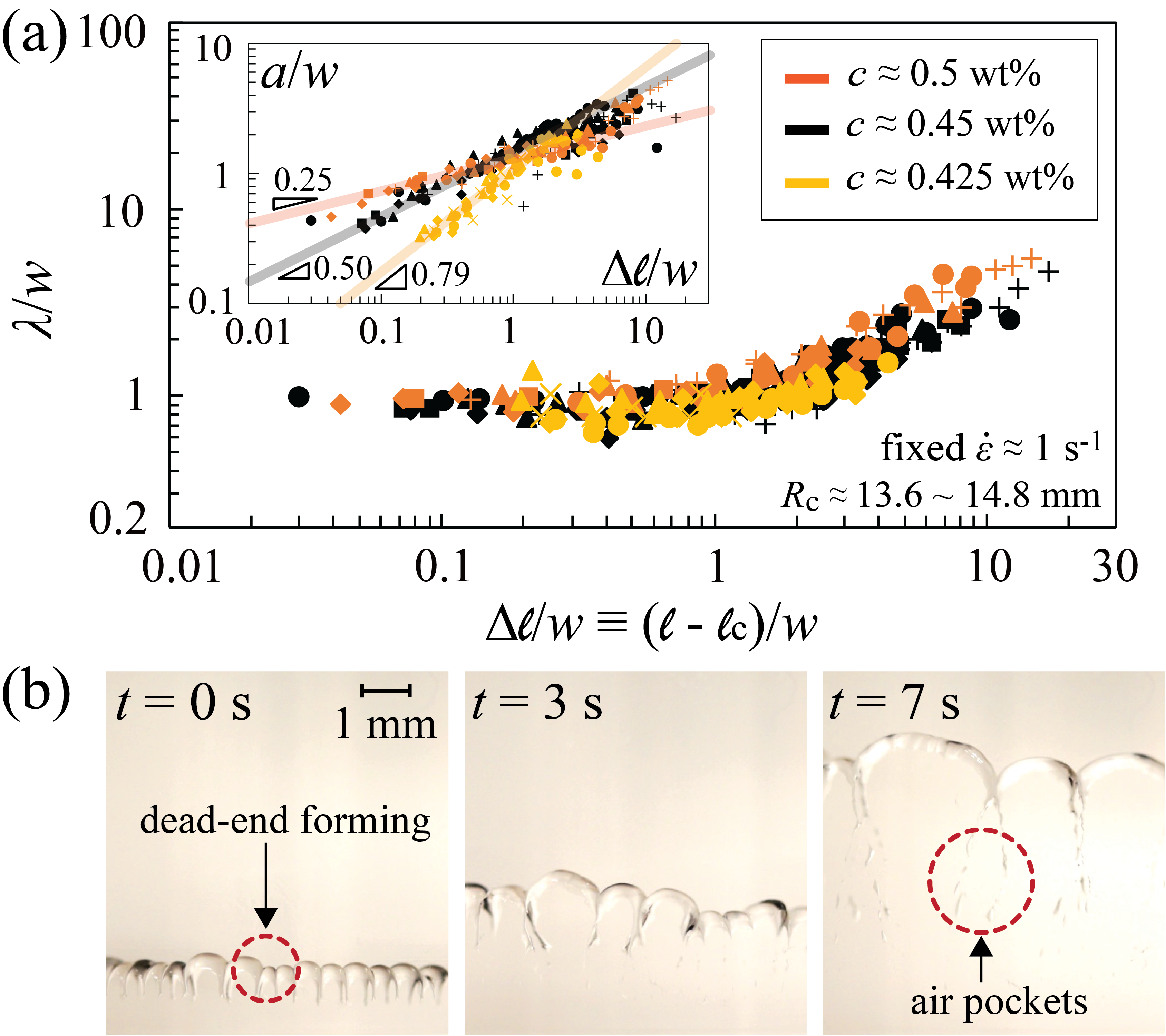}% Here is how to import EPS art
\caption{\label{fig: Figure4}Instability characteristics of continued extrusion. (a) $\lambda/w$ versus excess strain $\Delta\ell/w=(\ell-\ell_c)/w$ for varying $c$. Different symbols represent different values of $w/R_c$. For all $c$, $\lambda\sim w$ when $(\ell-\ell_c)<w$ and coalescence starts after $(\ell-\ell_c)\approx w$. (Inset) The furrow depth $a$ grows faster in softer gels. Lines are drawn with fitted power-law exponents. (b) Coalescing protuberances entrain air pockets behind (see Supporting video 1).}
\end{figure}

As the extrusion is continued, we measured the wavelength $\lambda$ and the furrow depth $a$ for different values of $c$ and $\dot\varepsilon$, as a function of the excess strain defined as $(\ell-\ell_c)/w$ [Fig.~\ref{fig: Figure4}; ref.~\cite{SupportingInformation}, \S~VI]. Regardless of $c$ and $\dot\varepsilon$, the furrows evolve longer wavelengths once $(\ell-\ell_c)>w$. The growth of furrow height $a(\ell(t))$ depended on the elasticity of gels [Fig.~\ref{fig: Figure4}(a); see Fig.~\ref{fig:Figure1}(b) to find $a$]. We observed no strong dependence of $a$ on $\dot\varepsilon$, which supports our conclusion that the instability is elasticity-driven. 

The observed phenomenon has significance for both industrial and biological contexts. For example, coalescence of the furrows engulfed trails of air pockets following behind the advancing gel front [Fig.~\ref{fig: Figure4}(b)]. Such air pockets can manifest as defects in objects manufactured through the extrusion of viscoelastic melts~\cite{gim2022review}. Additionally, the rheological characteristics of the soft solids investigated in this work bear qualitative resemblance to those of crosslinked networks found within cellular bodies. Consequently, our work may offer qualitative comparison between biological soft-solids and synthetic soft-solids moving inside confinement.

In conclusion, our study demonstrates the remarkable deformability of soft solids, showcasing their ability to flow into narrow gaps by continuously everting their bodies through shear interaction with the confining walls. Prolonged eversion of the thin annular soft body resulted in a novel form of surface instability, we termed the ‘furrowing’ instability. Our experiments showed that the critical strain and the instability wavelength depend only on the geometric length scales. These findings shed light on fundamental aspects of material behavior and also carry implications for both industrial processes and biological systems, offering insights into the motion and stability of soft bodies within confinement.

\begin{acknowledgments}
We thank Antonio Perazzo and Randy H. Ewoldt for advice on rheological measurements, Meisam Zaferani with the flow visualization, Glenn Northey with the setup, and Lihua Jin and John W. Hutchinson for discussions on the crease instability. We thank the Princeton Materials Research Science and Engineering Center (MRSEC, DMR-2011750) and the Kwanjeong Educational Foundation Fellowship for funding this work.
\end{acknowledgments}

%\nocite{*}

%\bibliography{References}% Produces the bibliography via BibTeX.

\end{document}

\preprint{APS/123-QED}

\title{Supporting Information for ``Surface Furrowing Instability \\ in Everting Soft Solids"}% Force line breaks with \\

\author{Jonghyun Hwang}
\author{Mariana Altomare}
\author{Howard A. Stone}
 \email{hastone@princeton.edu}
\affiliation{%
 Department of Mechanical and Aerospace Engineering, Princeton University, NJ 08540, USA
}%


\renewcommand{\figurename}{Fig.}
\renewcommand{\thefigure}{S\arabic{figure}}
\renewcommand{\theequation}{S\arabic{equation}}

\maketitle

\tableofcontents

\section{Microstructure and Rheology of Viscoelastic Solids}
\begin{figure*}
\includegraphics{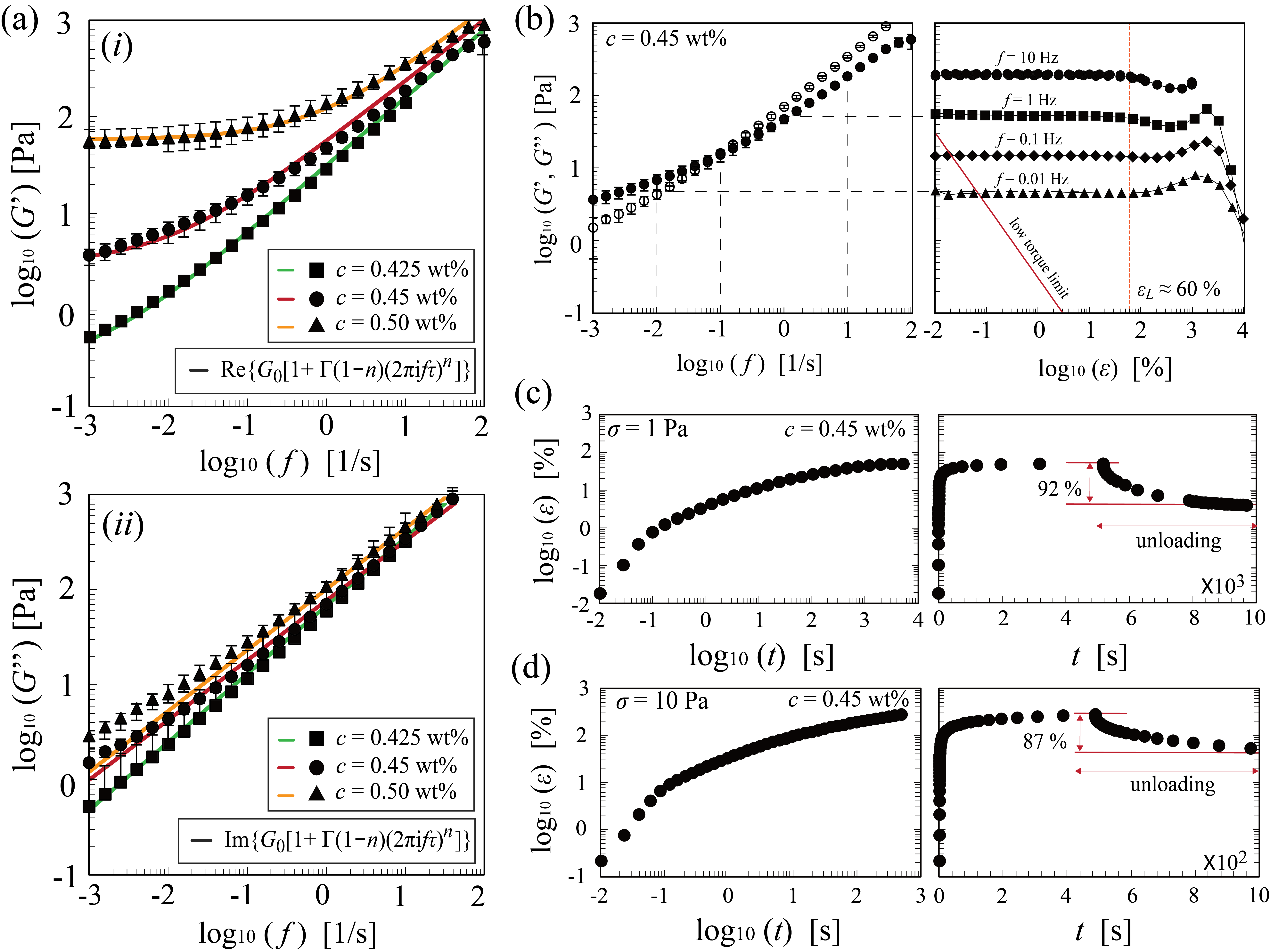}% Here is how to import EPS art
\caption{\label{fig:FigureS1}Rheological characterization of a soft viscoelastic solid. (a) Storage and loss moduli. (\textit{i}) The storage modulus is significantly affected by the small changes in the crosslinker concentration $c$. (\textit{ii}) The loss modulus is rather consistent for  different values of $c$. (b) Amplitude sweep measurements. (left) Storage and loss moduli at $c=0.45$ wt\%. (right) Corresponding measurements of $G'$ as a function of the shear amplitude at varying oscillation frequencies. The limit of linearity is observed to be $\varepsilon_L\approx60$ \%, for all oscillation frequencies tested. At frequencies higher than the crossover frequency, $f>f_c\approx0.1$ Hz, strain softening is observed owing to the relaxation of the side chains. The strain at maximum softening $\varepsilon_m\approx400$ \%, and the strain at fracture $\varepsilon_f\approx2500$ \%. (c) and (d) Measurements of creep (left panels) and creep and recovery (right panels) at applied shear stresses of 1 Pa and 10 Pa. We used $c=0.45$ wt\% gel. The gel recovers nearly 90 \% of its strain even after more than 1000 seconds of applied stress.}
\label{fig:S1}
\end{figure*}

We performed small amplitude shear oscillatory (SAOS) measurements to probe microstructural characteristics of our gel material. We used an Anton Parr MCR302e rheometer with a parallel plate geometry. The gel was cured overnight ($\geq$ 15 hours) on the rheometer plate at room temperature. The diameter of the plate was 50 mm, and the gap between the plates was maintained at 1 mm. We used sandblasted surfaces to prevent slippage of the material on the surface. The frequency sweep measurement was performed at a constant 1\% shear amplitude, from $f=0.001~\text{Hz}$ to $100~\text{Hz}$. From this measurement, we obtained the frequency dependent complex modulus $G^*(f)=G'+iG''$ [Fig. \ref{fig:FigureS1}(a)].
\\

\textbf{\textit{Theory of the loosely crosslinked network - }}Measurement suggested the existence of an overarching network that encloses the sol elements (silicone oil) at crosslinker concentration $c>0.4 ~\text{wt\%}$. In this state, the material is rubber-like, and $G'>G''$ in the low frequency limit. This rubber-like viscoelastic elastomer shows complex nonlinear relaxation phenomena, which can be modeled by considering relaxation process of dangling chains (side chains), such that $G_k(t)\sim\sigma_k(t)P(k)$~\cite{curro1983Macromol}, where  $k$ is the length of a dangling chain, $\sigma_k(t)$ is the stress that is assumed to grow proportional to the un-relaxed length of the side chain, $\sigma_k(t)\propto k-r(t)$ and $P(k)$ is the probability of finding a side chain of length $k$. The relaxed portion of chain $r(t)$ increases logarithmically in time \cite{DeGennes1979}. Assuming mean field percolation, $P(k)=q(1-q)^{k-1}$, where $q$ is the crosslink density normalized by the chain segment density. Summing the contribution to the stress over all possible chain lengths, one obtains,
\begin{equation}
G(t)=G_0[1+(\tau/t)^n]
\end{equation}
where the exponent $n$ is linearly proportional to  $q$, and $\tau\propto(aq^2)^{-b/q}$, where $a$ and $b$ are material parameters.

Dynamic response of the viscoelastic solids can be found by computing the transfer function of the above expression \cite{chaplain1968Nature}, which gives, 
\begin{equation}
G(s)=G_0[1+\Gamma(1-n)(s\tau)^n]
\end{equation}
where $\Gamma$ is the gamma function and $s$ is the transform variable $s=2\pi if$, where $i=\sqrt{-1}$. Frequently, a simplified form of dynamic modulus that omits the gamma function from the above expression is used, and yet provides adequate fit \cite{hwang2023langmuir}. Chasset and Thirion, who first empirically proposed this power-law relaxation function, noted the inadequacy of this expression to explain the long-time relaxation behavior (towards decreasing $f$, see Fig. \ref{fig:FigureS1}(a,\textit{ii}))  \cite{ChassetThirion}. From this expression, storage and loss moduli can be obtained be seeking the real and the imaginary part of $G(s)$, respectively. 
\\

\textbf{\textit{Stress and strain response of the viscoelastic solid gel - }}Next, we performed amplitude sweeps to determine strain-dependent behavior of the material. In Fig. \ref{fig:FigureS1}(b), we show the amplitude sweep result for a $c=0.45\text{ wt}$ \% gel probed at varying shear oscillatory frequencies. We observe that at frequencies less than the crossover frequency, $f<f_c\approx 0.1$ Hz, strain stiffening is observed until the shear strain reaches the fracture strain $\varepsilon_f\approx2500$ \%. For $f>f_c$, strain stiffening follows initial softening of  $G'$. This initial softening suggests strain-induced relaxation of the side chains. 

Lastly, we measured the change in strain at a constant applied stress for $c=0.45\text{ wt}$ \% gels. The time scale of the extrusion (inverse of strain rate $\dot\varepsilon$) ranges from 1 to 50 seconds. This is the time scale that a material element that enters the tip region takes to arrive at the confining walls, and also the duration of the applied stress on a material element near or at the free surface. We estimate the scale of the applied stress $\sigma$ within the tip region to be $\sigma\sim U_b/V=O(G_0)$, which is $O(0.1-10)~\text{Pa}$,  where the bending energy is $U_b\sim G_0w^2R_c$ and volume $V\sim w^2R_c$, as discussed in the main text. We chose two different values of shear stress, $\sigma = 1\text{ Pa}$ and $10\text{ Pa}$. The results show that the creep strain $\varepsilon(t)$ at time $t=\dot\varepsilon^{-1}$ for $\sigma=O(1)\text{ Pa}$ is $\varepsilon=O(10-100)$ \% and that after sufficient unloading period is $\varepsilon= O(1-10)$ \% [Fig. \ref{fig:FigureS1}(c)]. For both  values of the applied stress, the gel was able to recover approximately 90 \% of its strain after more than 1000 s of applied strain. Drawing a conclusion from this observation, we expect that the gel would recover almost its entire strain once the stress of $O(G_0)$ is removed after having been applied for $1 - 50$ seconds. Therefore, we consider that the material behavior is elastic for the timescales that we performed our experiments. 

\textbf{\textit{Comment on viscous dissipation - }} We argue that viscous effects are negligible. Based on the observations made in Fig. 3(b) in the main text, we observe that material flows vertically upward near the central region of the gap and flows vertically downward following the relative motion of the confining walls. Therefore, upward-flowing and downward-flowing material interact – these two regions are not directly crosslinked with each other; therefore, interactions mainly come from dangling chains and the Newtonian solvents. We justify this assumption based on the composition of the material – a loose elastic network filled with low viscosity silicone oil. The timescale of the viscous interaction $\tau_{viscous}=\eta/G_0= O(10^{-2} ~\text{Pa}\cdot\text{s})/O(10^{-1}-10 ~\text{Pa})= O(10^{-3}~\text{s})$ for stiff gels to $= O(10^{-1}~\text{s})$ for soft gels. On the other hand, the time scale of the relaxation of the dangling chains $\tau_{elastic}$, which we define as the inverse of the $G'$ and $G''$ crossover frequency, is $ O(10^{-1}~\text{s})$ for stiff gels to $= O(10^{3}~\text{s})$ for soft gels. Also, the time scale of the extrusion $\tau_{extrusion}$,  which we define as the inverse of the shear rate, is $O(10^{0}~\text{s})$ to $ O(10^{2}~\text{s})$. Therefore, $\tau_{viscous}\ll \tau_{elastic} \sim \tau_{extrusion}$. Hence, we treat that elastic interaction as most relevant for the time scale of our experiment.

\textbf{\textit{Independence of $\ell_c$ and $\lambda$ on $c$ or $G_0$ - }} In Fig. 2. of the main text, we show that $\ell_c$ and $\lambda$ are independent of how cross-linked the gel is (thus $c$ and $G_0$) and of how fast the gel is being extruded into the annulus ($\dot\varepsilon$). The explanation to the latter observation is physically intuitive as the response of the gel to the imposed stress, although being viscoelastic due to having dangling chains, is most strongly governed by the elasticity of the ``infinite network'' \cite{rubinstein2003polymer}. We still note that, once the extrusion is stopped, the buckled free surface slowly relaxes in amplitude to some extent (due to the time-dependent relaxation of the dangling chains) but maintains the overall shape. Although we did not quantify this observation, the time-scale of relaxation was qualitatively consistent with the relaxation time measured from the rheological measurement.

Next, the observation made with regards to the independence of $\ell_c$ on $G_0$ was consistent in the range of $G_0$ we explored in this study ($G_0=0.2\sim 55$ Pa). Below $c=0.425$ wt\%, the material stayed below the gel point (thus having no ``infinite network'') and there was no furrowing instability in the range of $\dot\varepsilon$ we explored in this study. And above $c=0.55$ wt\%, the gel was too stiff to be extruded into the channel. We concluded that the key factor to observe the furrowing instability is the existence of the infinite network with high fraction of loose dangling chains.

\section{Periodicity of the Furrowing Instability}
\begin{figure*}
\includegraphics{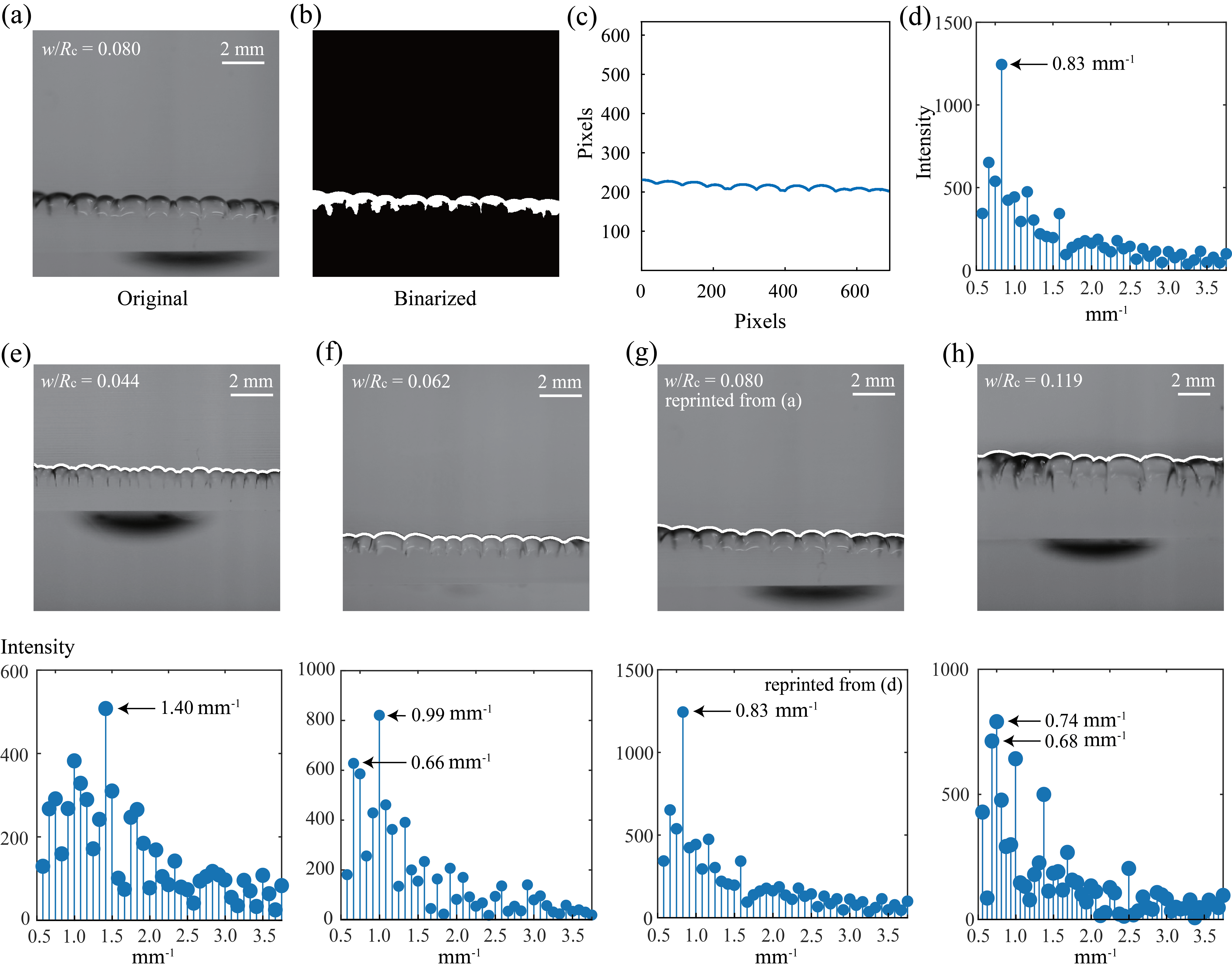}% Here is how to import EPS art
\caption{\label{fig:FigureS2}Periodicity of the instability. (a) An original side view image depicting a chord length of 12.1 mm. (b) Binarized image. (c) Traced interface to obtain a 1D data of the deformed free surface. (d) Fast Fourier Transform (FFT) is applied to the 1D data in (c), and the strongest ‘per length’ frequency is obtained. In this case, 0.83 mm\textsuperscript{-1} at the strongest peak. This value corresponds to 10 periodically spaced beads, which agrees with the visual shown in (a). (e) - (f) The same analysis with varying $w/R_c$; all data show peak(s) in the wavelength, confirming the periodicity of the observed instability.
}
\end{figure*}

We performed image analysis on the snapshots of the side view videos to comment on the periodicity/aperiodicity of the surface instability. We used a Nikon D3300 with a 200 mm focal length lens; the camera imaged a segment of 12.1 - 16.1 mm. The ratio between the length of the segment (\textit{c}) to the actual arclength (\textit{a}) of the material in the curved geometry was 1:1.03 - 1.05. By using MATLAB Fast Fourier Transform feature, the periodicity of the initial wavelength was analyzed, and the method is described in Fig. \ref{fig:FigureS2}. In the example shown in  Fig. \ref{fig:FigureS2}(a-d), the strongest frequency signal occurs at 0.83 mm\textsuperscript{-1}. This suggests the number of the feature to be 0.83 mm\textsuperscript{-1} $\times$ 12.1 mm = 10, which agrees with our visual observation of $n\approx10$. From Fig. \ref{fig:FigureS2}(e) through (h), we show four other examples with varied $w/R_c$.

\section{Furrowing Instability in Other Geometries}
\begin{figure*}
\includegraphics{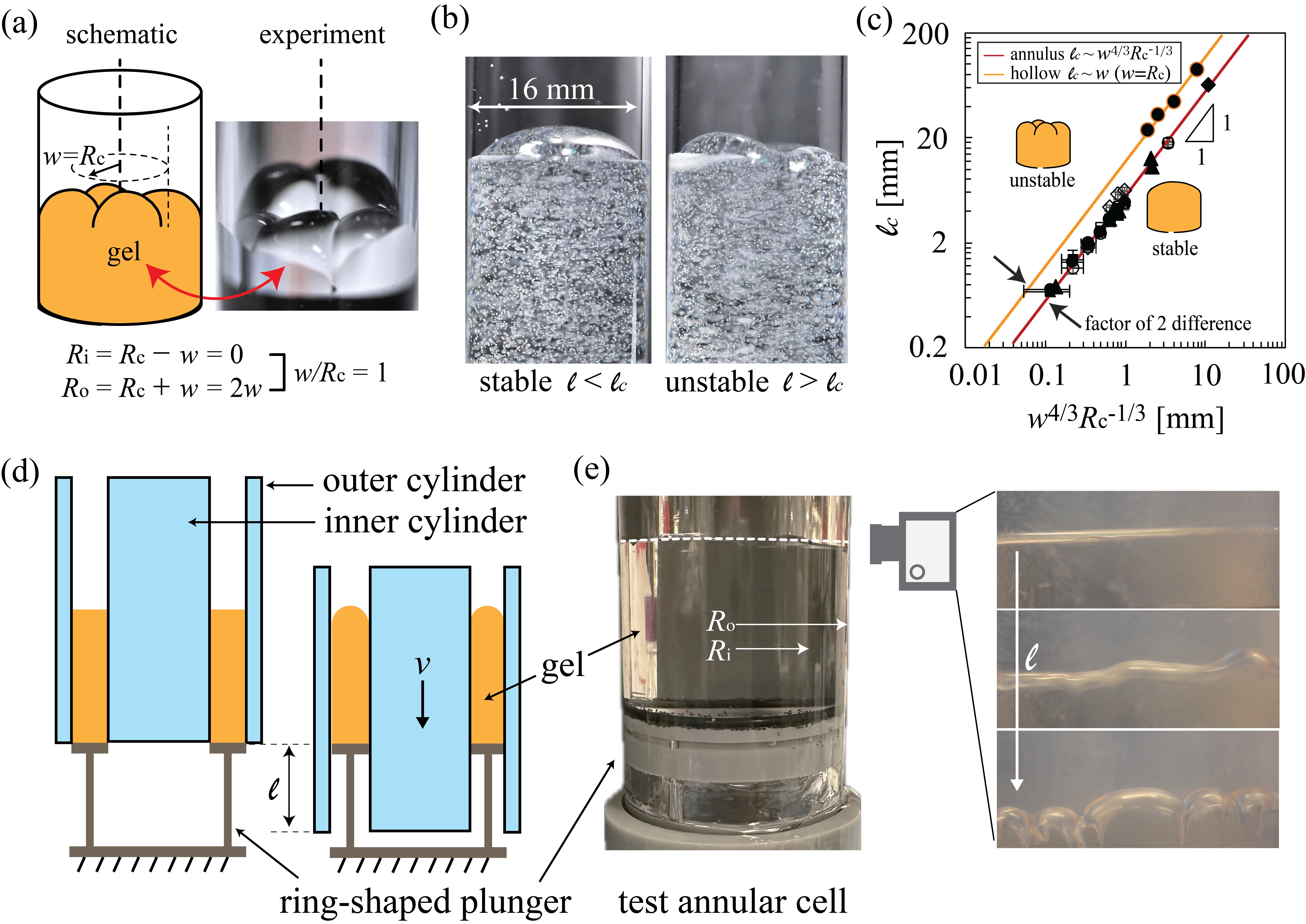}% Here is how to import EPS art
\caption{\label{fig:FigureS3}Furrowing instability in the absence of gel reservoir. (a-c), Hollow cylinder experiment. (d-e), ‘Without reservoir’ annulus experiment. (a) Geometrical interpretation of the hollow cylinder geometry as an annulus with vanishing radius of the inner cylinder $R_i=R_c-w\rightarrow0$. The free surface of the gel deforms to show a flower-like morphology (see angled view). (b) Side view of the gel before and after the onset of the furrowing instability. (c) Measurement of $\ell_c$ overlaid on top of the plot from Fig. 2(a) for comparison. Measurement shows $\ell_c\sim w$ but has a different prefactor compared to the measurements taken with the annular geometry. (d) Schematic of the ‘without reservoir’ experiment. Recall that, in the main text, we presented a geometry comprised of an ‘annulus’ and ‘reservoir’ [Fig. 1(a)]. (e) We observe that the furrowing instability forms due to shear interaction at the wall and without any overall solid-body compression.
}
\end{figure*}

In the main text, we presented experiments performed with annular geometries with the gel reservoir (see Fig. 1(a) in the main text). In this note, we present a couple of model experiments where we observe the furrowing instability in other geometries. In Fig. \ref{fig:FigureS3}(a-c), we present experiments performed with hollow cylinders of various radii. In Fig. \ref{fig:FigureS3}(d-e), we demonstrate the formation of the furrowing instability without the gel reservoir. In both experiments, there is no change in the nominal area (normal to the axial direction) in which the gel is confined during the extrusion of the gel.

To analyze the data, we treat the hollow cylinder geometry as an annulus with its inner radius vtending to zero, $R_i=R_c-w\rightarrow0$. Since $w$ is defined as half of the gap between the two confining walls, \textit{w} is then the radius of the hollow cylinder. Thus, we obtain $\ell_c\sim w^{4/3}R_c^{-1/3}\sim w$ [Fig. \ref{fig:FigureS3}(a)]. Therefore, we hypothesize that the free surface of the gel confined inside a hollow cylinder would become unstable once extruded for $\ell_c\sim w$, as discussed in the main text. An example is shown in Fig. \ref{fig:FigureS3}(b). We performed experiments using \textit{c} = 0.45 wt\% gel extruded at $\dot\varepsilon=0.1\text{ s}^{-1}$. The results obtained from the `hollow cylinder' experiments are overlaid on top of the `annulus with reservoir' experiments in Fig. \ref{fig:FigureS3}(c). The result confirms that the furrowing instability is not specific to the annular geometry. The two sets of measurements have differences of a factor of 2, which we believes stems from the geometry changing from an annulus to a confined cylindrical channel, as illustrated in Fig. \ref{fig:FigureS3}(a). We also note that the perturbation did not start from the tip of the curved front, but rather initiated from the confining wall, which is consistent with the observation made in the main text.

Next, we describe the `without reservoir' experiment. In our earlier experiments (described in Fig. 1(a) of the main text), we conducted tests using a setup comprising the `annulus' and `reservoir' sections. Gels entering the annulus from the reservoir may experience radial and circumferential compression. To ascertain that this compression is not the cause of the furrowing instability observed, we designed the `without reservoir' experiment. We prepared an annulus of outer diameter 31 mm and inner diameter 25 mm. Then, we cast a $c=0.45$ wt\% gel in the gap of the annulus. The gel was cast on top of a ring-shaped plunger and left to cure  overnight. Once the gel was fully cured, we pushed the annular cell casing the gel straight downward at a constant strain rate of  $\dot\varepsilon=0.1\text{ s}^{-1}$, as illustrated in Fig. \ref{fig:FigureS4}. Therefore, the ring-shaped plunger travels upward relative to the motion of the casing. The gel interface was initially flat; however, after some length $\ell$, the free surface of the gel deformed and showed the furrowing instability. 

From the `hollow cylinder' experiments and the annulus `without reservoir' experiment, we conclude that the furrowing instability is not caused by the solid body compression, but rather appears whenever `eversion' of the solid body occurs.

\section{Comparison to Compression-driven Wrinkling/Crease Instability}
\textbf{\textit{Justification for a  linear stability analysis - }}There is geometric similarity between the phenomenon considered in this article and the so-called ‘creasing’ instability \cite{ghatak2007PRL,hohlfeld2011PRL}. The creasing instability is a subcritical bifurcation of the Biot instability, which is the formation of localized surface wrinkles in a laterally compressed elastic block. Since creases form at a lower critical compressive strain compared to wrinkles, observation of the Biot instability had been a challenge. However, works of Mora et. al. \cite{mora2011SoftMat} and Liu et al. \cite{liu2019PRL} showed that in presence of strong surface interactions (increasing $\gamma/G_0 L$, where $L$ is the characteristic length scale of the problem), the critical compressive strains of the two types of surface instability become alike. The work of Cao and Hutchinson characterize the creases as the ‘collapsed’ state of the wrinkle \cite{cao2012RSPA}. This suggests that one can perform linear stability analysis to identify the critical condition for the formation of the surface instability, when the material experiences a strong surface effect. 

In this section, we will delineate the effect of squeeze compression on the formation of the surface instability by following the work of Cai et. al. \cite{cai2010SoftMat} and Liu  et. al \cite{liu2019PRL}.
\\

\begin{figure*}
\includegraphics[width=0.9\linewidth]{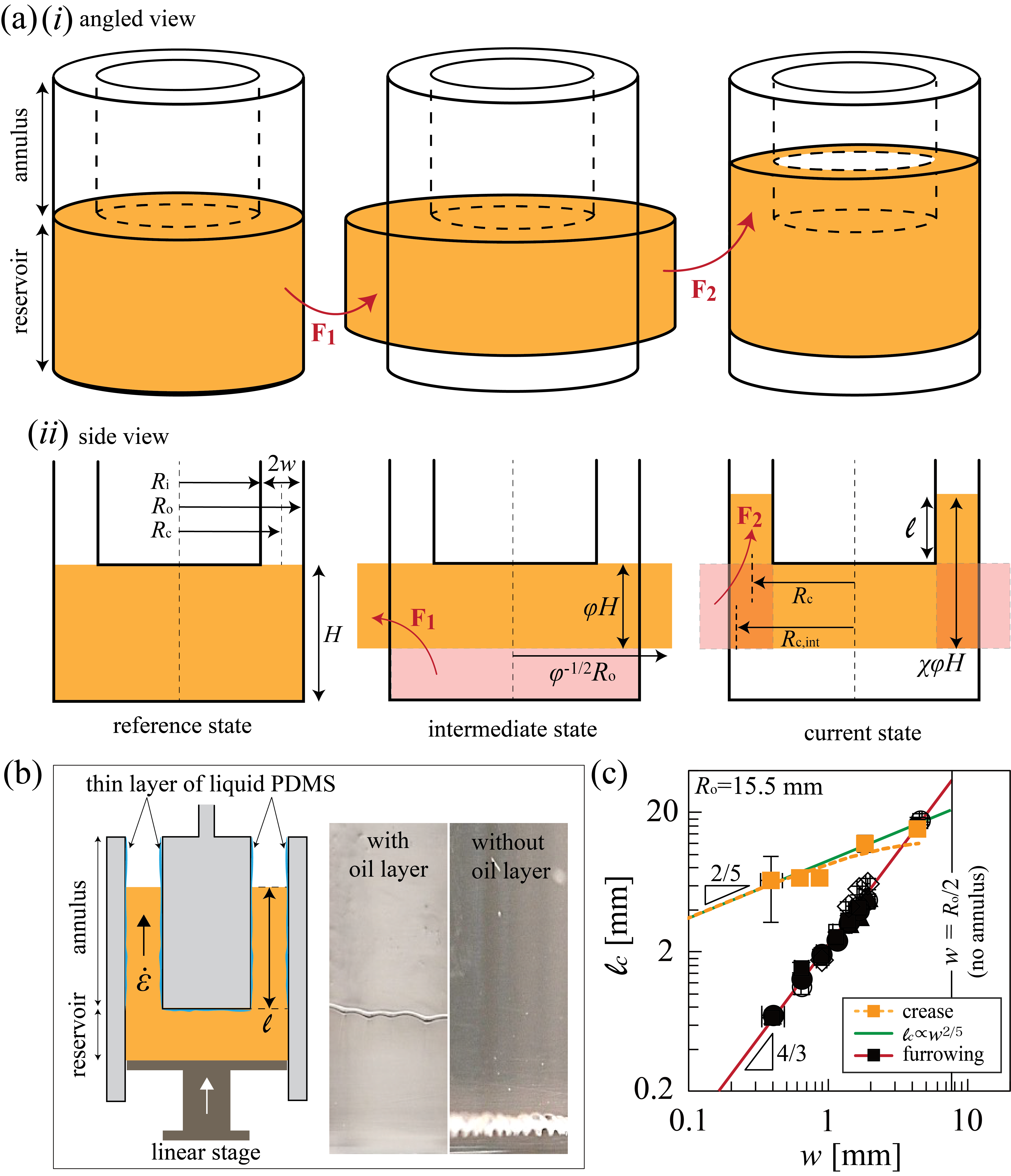}% Here is how to import EPS art
\caption{\label{fig:FigureS4}Schematics of the deformation. We assume a thin-ring annulus limit $w\ll R_c$ such that deformation is quasi two-dimensional. (a) View at an angle (\textit{i}), and the side view (\textit{ii}). The gel in the reservoir in the reference state is stress free. The gel is squeezed vertically with a compression ratio of $\varphi$ from the reference state to the intermediate state, through the deformation gradient $\textbf{F}_1$. Next, the squeezed gel portion between $\varphi^{-1/2}R_o>r>R_i$ is then deformed to the current state to fit inside the annulus by the deformation gradient $\textbf{F}_2$. (b) Schematics of the design of experiment. We wet the surface of the annulus with a thin layer of silicone oil. The gel slips on the surface of the wall and the shear interaction is suppressed. (c) We perform the linear stability analysis and obtain the critical length $\ell_c$ for the compression driven instability (yellow dashed line). The theory is valid only at small $w$. We overlay $\ell_c\propto w^{2/5}$ line in blue to highlight asymptotic behavior of the crease instability in the thin-ring annulus limit. Measurements from (b) agree with the calculated crease theory line, showing the distinction from the furrowing instability (red curve and black markers). }
\end{figure*}

\textbf{\textit{Modeling solid body deformation - }}We start by describing the elastic deformation of a squeezed gel body, as described in Fig. \ref{fig:FigureS4}. In Fig. \ref{fig:FigureS4}(a, b), we show a set of schematics that describe the deformation of the gel in the ‘with reservoir’ case. The gel is compressed in the vertical direction, and then the gel elements squeezed outside $R>R_i$ further deform to fit inside the annulus. Thus, we decompose the deformation in two steps. Typically, the total deformation gradient is obtained by multiplying the deformation gradients in different steps. However, since we assume inhomogeneous deformation for the second step in the deformation, we set the total deformation gradient to be the composition of the independent deformations, $\text{\textbf{F}}=\text{\textbf{F}}_2(\text{\textbf{F}}_1)$. 

Because the gel is incompressible, the radial and azimuthal components of the deformation gradient are chosen to satisfy det($\textbf{F}_1$) = 1.  Assuming a homogeneous axisymmetric deformation, i.e., the deformation is the same throughout the sample, the first step in the deformation can be described with $\text{\textbf{F}}_1=\text{diag}(\varphi^{-1/2},\varphi^{-1/2},\varphi)$, where $\varphi$ is the final length divided by the initial length, hence a measure of stretch. This changes the height of the gel to be $\Delta h=(\varphi-1)H$. Through this deformation, material is squeezed and moves homogeneously from the center  radially outward. Next, the second step in the deformation \textbf{F}\textsubscript{2} describes how the excess portion of the squeezed material ($r>R_i$) would deform to fit within the annular section. During the second step of deformation, the inner boundary of the annulus is prescribed such that the squeezed material in $r<R_i$ does not deform. We will write \textbf{F}\textsubscript{2 }with a set of arbitrary stretches, $\text{\textbf{F}}_2=\text{diag}(\chi_r,\chi_\theta,\chi_z)$ and express each component as a function of geometry and the squeeze compression $\varphi$.

 In the case that the annular curvature $\kappa=w/R_c$ is small, one can assume the deformation in the radial direction is homogeneous. With this assumption, the material in $[(R_c-w),\varphi^{-1/2}(R+w)]$ will be homogeneously deformed to $[(R_c-w),(R+w)]$ [Fig.\ref{fig:FigureS4}(a)]. The radial compression associated with this deformation is then
\begin{equation}
 \chi_r=2\kappa/[\varphi^{-1/2}(1+\kappa)-(1-\kappa)]. 
\end{equation}
 Next, azimuthal compression is the change in the radius of the centerline component of the gel before and after the deformation. Specifically, the centerline radius of the squeezed portion of the gel (in the intermediate state [see the middle panel in Fig. \ref{fig:FigureS4}]]) is $R_{c,int}=R_i+(\varphi^{-1/2}R_0-R_i)/2$. This line element is deformed to $R_c=R_i+w$ in the current state. Then, the corresponding compression ratio in the azimuthal direction is 
\begin{equation}
 \chi_\theta=R_c/R_{c,int}=2/[\varphi^{-1/2}(1+\kappa)+(1-\kappa)].
\end{equation} 
To enforce incompressibility, 
 \begin{equation}
 \chi_z=(\chi_r\chi_\theta)^{-1}=[\varphi^{-1}(1+\kappa)^2-(1-\kappa)^2]/(4\kappa).
 \end{equation}
 
Now, we will relate the arbitrary compression ratio $\varphi$ to the physical variables in the system. First, the displaced height of the material equals $(1-\varphi)H=v_{pump}t$, where $v_{pump}$ is the vertical speed of pump. Because $v_{pump}$, and the rise velocity of the interface, $v_{interface}$, are related by $\pi(R_c+w)^2v_{pump}=\pi[(R_c+w)^2-(R_c-w)^2]v_{interface}$, the following relationship holds: $(1-\varphi)H=4(v_{interface}t)\kappa/(1+\kappa)^2$. To distinguish the effect of compression from the wall friction, we will assume that the gel slips on the surface of the wall (at $r=R_i$ and $R_o$). Then, one obtains height of the gel reservoir to be $H=4(v_{interface}t)\kappa/[(1+\kappa)^2(1-\varphi)]$. Also, because $v_{interface}t=\ell=(\chi_z-1)\varphi H$, we obtain that $\chi_z=1+(1-\varphi)(1+\kappa)^2/(4\varphi H)=[\varphi^{-1}(1+\kappa)^2-(1-\kappa^2)]/(4\kappa)$.
This agrees with the expression of $\chi_z$ that we obtained in the preceding paragraph. Finally, since $H=v_{pump}t_f$, where $t_f$ is the time taken to fully extrude the gel, $(1-\varphi)=t/t_f$. Also, since $\ell_f=v_{interface}t_f$, then $t/t_f=\ell/\ell_f$. Therefore, 
\begin{equation}
\varphi=1-\ell/\ell_f.
\end{equation}

\textbf{\textit{Linear perturbation analysis - }}We start by identifying the governing set of equations. Assuming an ideal elastic body, we employ a neo-Hookean strain energy function $W$ as 
\begin{equation}
W=\frac{1}{2}G_0(F_{iK}F_{iK}-3)-p(\text{det}(\textbf{F})-1).
\label{eq:one}
\end{equation}

Here, \textbf{F} is the deformation gradient, $G_0$ is the shear modulus, and $p$ is the Lagrangian multiplier, and any equilibrium process during the solid body deformation involves minimization of this energy. The components in \textbf{F} are still allowed to vary independently. The true stress is given as
\begin{equation}
\sigma_{ij}=\frac{F_{jK}}{\text{det}(\textbf{F})} \frac{\partial W(\textbf{F})}{\partial F_{iK}}-p\delta_{ij},
\end{equation}
and the perturbed stress is written as
\begin{equation}
\delta\sigma_{ij}=G_0F_{jK}F_{pK}\delta f_{ip} - \delta p(X)\delta_{ij} + p(X)\delta f_{ji}.
\end{equation}
Here, $\delta f_{ji}$ is the perturbed displacement gradient in the current configuration (Eulerian framework). $\delta \textbf{f}$ and the perturbed deformation gradient $\delta\textbf{F}$ are related via $\textbf{M}=[\textbf{F}^{-1}]^\text{T}$ as

\begin{equation}
\delta f_{ij}=\frac{\partial\delta x_i(\textbf{X})}{\partial x_j}=\frac{\partial\delta x_i(\textbf{X})}{\partial X_K}\frac{\partial X_K}{\partial x_j}=\delta F_{iK}M_{jK},
\end{equation}

where $\delta \textbf{x}$ is the displacement in the current coordinates, $\textbf{x}$ is the position in the current coordinates, and $\textbf{X}$ is the position in the reference coordinates. The condition for incompressible deformation is
\begin{equation} \label{perturbed incompressibility}
\delta f_{ii}=\frac{\partial\delta x_i}{\partial x_i}=0 \text{ or } \nabla\cdot\delta \textbf{x}=0,
\end{equation}
and the mechanical equilibrium is achieved with
\begin{equation} \label{perturbed equilibrium}
\frac{\partial\delta\sigma_{ij}}{\partial x_j}=\textbf{0} \text{ or } \nabla\cdot\delta \boldsymbol{\sigma}=\textbf{0}.
\end{equation}
To perform the stability analysis, we use the following ansatz,
\begin{subequations} \label{perturbation ansatz}
      \begin{equation} 
   \delta x_\theta=a\text{e}^{\xi z}\text{sin}(n\theta),
     \end{equation}
\begin{equation} 
\delta x_z=b\text{e}^{\xi z}\text{cos}(n\theta),
\end{equation}
\begin{equation} 
\delta p=c\text{e}^{\xi z}\text{cos}(n\theta),
\end{equation}
\end{subequations}
where $n=2\pi R_c/\lambda$ is the wave number. Assuming an annulus in the shape of thin-ring ($w\ll R_c$) and perfect-slip at the wall, $\delta x_r$ and $\partial/\partial r$ are neglected. At the free surface ($z=0$), we assume no shear traction, $\delta\sigma_{\theta z}=0$. Along the normal direction, we then apply a stress boundary condition balancing elastic stress and capillary stress.
\begin{equation}
\hat{\textbf{n}}\cdot\delta\pmb{\sigma}_{air}\cdot\hat{\textbf{n}}-\hat{\textbf{n}}\cdot\delta\pmb{\sigma}\cdot\hat{\textbf{n}}=\gamma(\nabla\cdot\hat{\textbf{n}}),
\end{equation}
where $\hat{\textbf{n}}$ is the outward normal vector. We assume no shear stress from air, hence $\textbf{t}\cdot\delta{\pmb\sigma}_{air}=\textbf{0}$, where $\textbf{t}$ is the unit tangent vector. We also assume perturbation to vanish at a depth $h_0$, i.e., $\delta x_z(z=-h_0)=\delta x_\theta(z=-h_0)=0$.

To close the problem, we specify the Lagrangian multiplier, which is defined in the reference configuration, such that $p(\textbf{X})$. The solid body is stress-free in its reference configuration. In terms of nominal stress $\textbf{s}$, $s_{iK}=G_0F_{iK}-pM_{iK}$. Taking the vertical, stress-free component, $s_{33}=G_0\chi_z-p\chi_z^{-1}=0$, giving $p=G_0\chi_z^2$.

The linear perturbation analysis is done by seeking a nontrivial solution to the set of governing equations (Eqn. \ref{perturbed incompressibility} and \ref{perturbed equilibrium}) that satisfies the boundary conditions given above, using the assumed form of the perturbation ansatz (Eqn. \ref{perturbation ansatz}). One obtains the dispersion relation that contains three dimensionless parameters, which are $\gamma/wG_0, w/R_c$, and $h_0/w$. A most unstable stretch $\varphi=1-\ell_c/\ell_f$ is found as a function of these three dimensionless groups, which yields the ‘crease’ line in Fig.\ref{fig:FigureS4}(c). The only fitting parameter, $h_0$, which stands for the depth below the free surface where the perturbation vanishes does not significantly affect $\ell_c$, but did affect the most unstable wave number $n$. In calculations, we set $h_0=(1-\varphi)\ell_f$. We fitted the numerical results to obtain the approximation $\ell_c\propto w^{2/5}$ (blue line in Fig.\ref{fig:FigureS4}(c)). We emphasize again that our solution applies to the low $w/R_c$ limit with a slip boundary condition, where the deformation can be considered two dimensional. Towards the other limit, $w/R_c\rightarrow 1$, there would be no annulus, and hence the theory breaks down ($\varphi$ is constant 1 in this limit, and therefore there would be no solid body deformation).

\textbf{\textit{Comparison with experiment - }}In our linear stability theory, to highlight the role of compression, we assumed that the effect of wall shear is absent, i.e., a slip boundary condition. Recall that the crease instability occurs due to the global compression of a solid body, and the furrowing instability described in the main text occurs, we hypothesize, due to the forced eversion of the material due to wall shear. Therefore, in the geometry that includes the gel reservoir, we expect to observe the crease instability if the material-wall interaction is suppressed. This, of course, only applies to the annular geometry with the gel reservoir; for the geometries explored in Sec. III of this SI, we do not expect a creasing instability as there is no overall solid body compression. 

To suppress the no-slip boundary condition in the experiments, we wet the surfaces of the annulus with liquid PDMS. We chose silicone oil of 5 mPa$\cdot$s (Sigma Aldrich, used as purchased) (see Fig. S4 (b)). A thin layer of oil allowed fully cured gel to slip on the surface of the annulus, suppressing the wall shear. As expected, the instability formed at a higher $\ell_c$, agreeing with the calculation made in the preceding section (see Fig. \ref{fig:FigureS4}(c), yellow symbols). In all experiments, we used $R_o=15.5$ mm. We stress again that the model is valid in the limit where $w$ is small compared to $R_c$. This experimental observation confirms that furrowing instability is a distinct mode of surface instability from the crease instability.

\section{Curvature scaling and other supplements to the scaling argument}
We relate the height of the curved tip region $h$ to the geometrical length scale $w$. If one assumes that material at either side of the confining walls does not slip, then we can model the progressing gel ‘column’ as a one-dimensional structure where the addition of the new material is localized at the tip. We call such a localized region the ‘growth’ region. If the material slips on the wall and the addition of the new material can occur anywhere along the height of the entire column, such that $\frac{\partial \varepsilon}{\partial t} = k\varepsilon$, where $k$ is the rate of stretch, then the stretch of the new interface experiences exponential growth $\varepsilon \sim \text{exp}(kt)$, and the height of the column relative to the initial height $\ell_0$ would be $\ell=\ell_0(1+\varepsilon)=\ell_0+\ell_0 \text{exp}(kt)$. However, we do not observe exponential progression of the curved tip. 

\begin{figure*}
\includegraphics[width=1\linewidth]{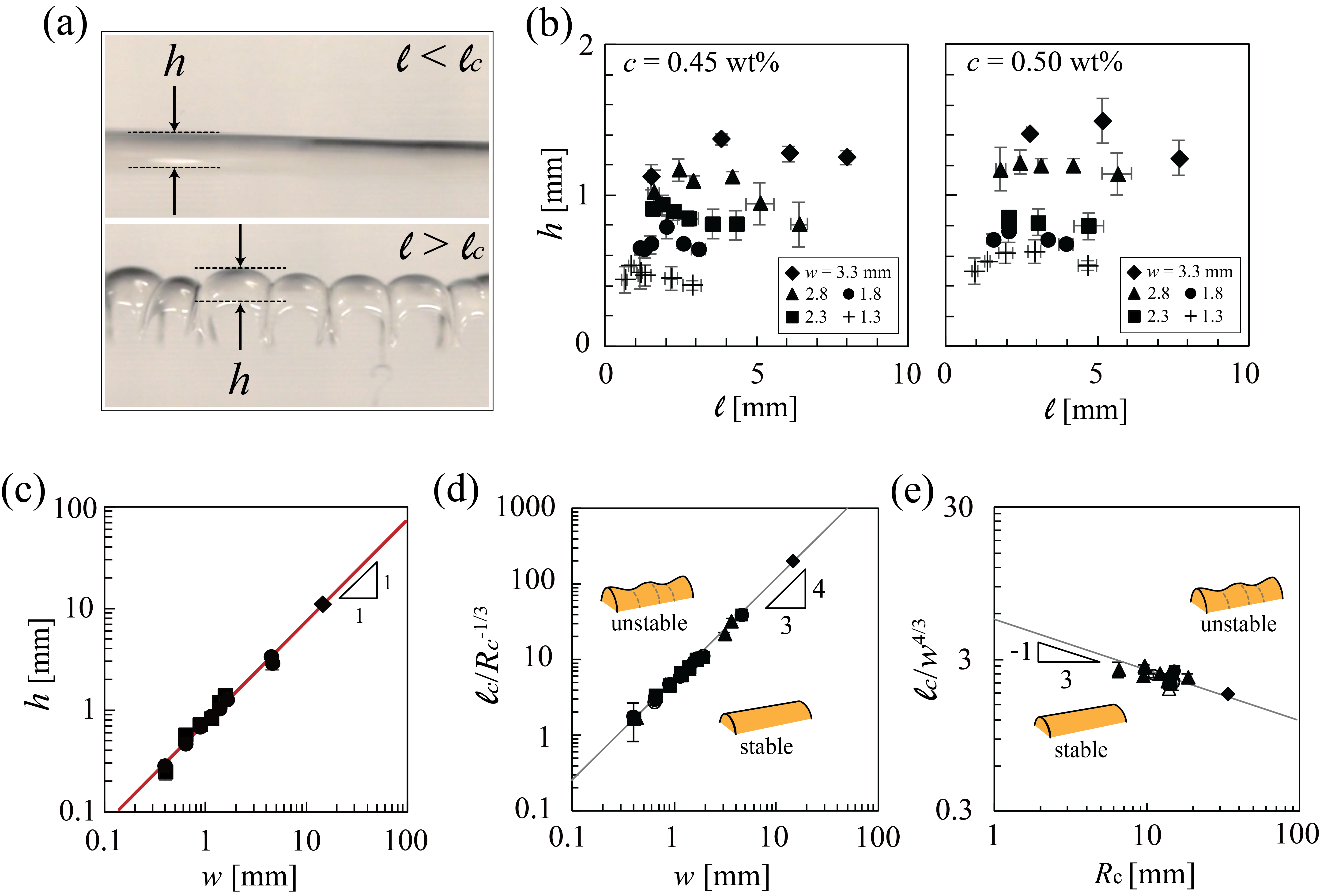}% Here is how to import EPS art
\caption{\label{fig:FigureS5}Measurement of the tip height $h$ as a function of $\ell$ and $w$. (a) We define $h$ as distance between the tip of the curved front region and the point where the meridian meets the confining wall. The same definition applies to post-instability ($\ell>\ell_c$) measurements. (b) Collections of representative measurements (not all) of $h$ as a function of $\ell$ for $c=0.45$ wt\% and $c=0.50$ wt\%. We observe that $h$ is nearly constant over $\ell$. We take average of each $h(w)$ over $\ell$ and plot the result in (c). (c). Collection of all data points of varied $c$ and $\dot\varepsilon$. We observe that $h\approx 0.75 w$. (d) Experimental measurements showing $\ell_c/R_c^{-1/3}\propto w^{4/3}$ and (e) measurements showing $\ell_c/w^{4/3}\propto R_c^{-1/3}$. Different symbols stand for different combinations of $w, R_0, \dot\varepsilon$, and $c$.}
\end{figure*}

If the material does not slip on the wall, the stretch as a function of time can be described as
\begin{equation}
\frac{\partial \varepsilon }{\partial t}=k\varepsilon (1-H(s-h))=
    \begin{cases}
      k\varepsilon & 0<s<h\\
      0 & h<s<\ell(t)
    \end{cases}   
\end{equation}
where $s$ is the coordinate along the height of the column, $h$ is the constant height of the growth region, and $H$ is the Heaviside step function. For convenience, we set $s=0$ at the free end of the growth region. This function cannot be directly integrated as \textit{s} also changes over time; $s=s(s_0,t)$. Hence, we consider a coupled system of ODE, 
\begin{equation}
    \begin{cases}
      \frac{\partial \varepsilon}{\partial t} &= k\varepsilon (1-H(s-h))\\
      \frac{\partial s}{\partial s_0} &= \varepsilon(t)
    \end{cases} 
\end{equation}
with initial conditions $\varepsilon|_{t=0}=1$ and $s|_{t=0}=s_0$ we obtain that 
change in height $\Delta\ell(t)=\ell(t)-\ell_0=kht$ \cite{goriely2017Book}. The initial height of the column $\ell_0$ is can be any value larger than $h$. Since $\Delta\ell$ changes in time by a constant average velocity $v=\dot\varepsilon w$, $kht=\dot\varepsilon wt$. Comparing the dimensions, we obtain that rate of growth $k\sim\dot\varepsilon$ and height of the growth region $h\sim w$. From experiments, we see that $h\approx0.75w$, independent of $\dot\varepsilon$ and $c$ [Fig. \ref{fig:FigureS5}].

Additionally, to further support our observation on the power-law response of $\ell_c$ shown on the main text, we have performed measurements of $\ell_c$ with varying confinement scale $w$ and $R_c$ in Fig.~S5(d) and (e). The result shows that $\ell_c\propto w^{4/3}$ and $\ell_c\propto R_c^{-1/3}$, respectively.

\begin{figure*}
\includegraphics[width=1\linewidth]{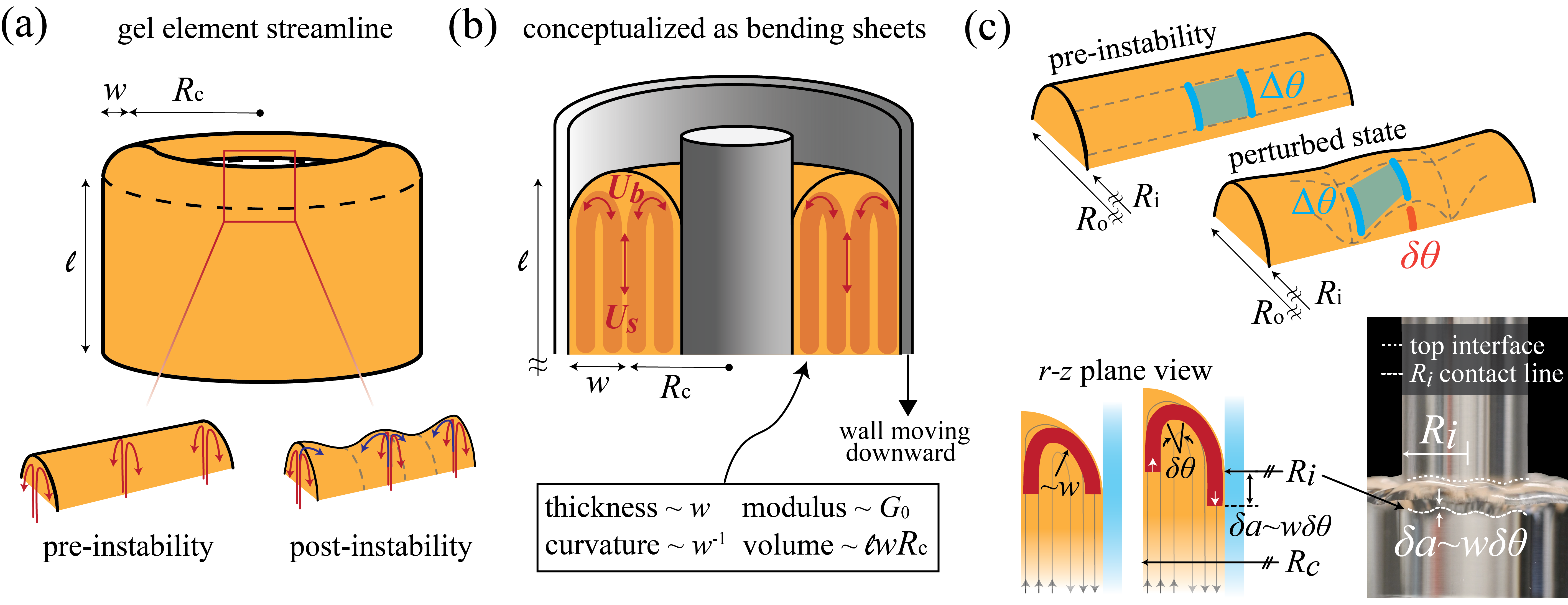}% Here is how to import EPS art
\caption{\label{fig:FigureS6}Schematics to supplement ideas used in the scaling argument. (a) streamline of gel elements in co-moving frame of reference to supplement experimental observation made in Fig 3(a) and (b) in the main text. (b) Conceptualization of the rotating gel streamlines as bending elastic sheets, where bending only occurs within the tip region of curvature $\kappa\sim w^{-1}$. (c) Once immediately neighboring gel elements (colored in blue) are displaced by $\delta\theta$ due to perturbation. This generates torsional energy of $\delta U_t$.}
\end{figure*}

In Fig. \ref{fig:FigureS6}(a), we provide schematic representations of the gel streamlines in the co-moving frame of reference (see schematic in Fig. 3(b) in the main text) as a supplement to the experimental observation made in Fig. 3(a) and (b) in the main text. Also, we illustrate the conceptualization of rotating gel vortices as bending elastic sheets in Fig. \ref{fig:FigureS6}(b), and and indicate how the angular perturbation $\delta\theta$ gives rise to torsional energy in Fig. \ref{fig:FigureS6}(c)

\section{Instability wavelength and furrow depth at prolonged extrusion}
\begin{figure*}
\includegraphics[width=0.9\linewidth]{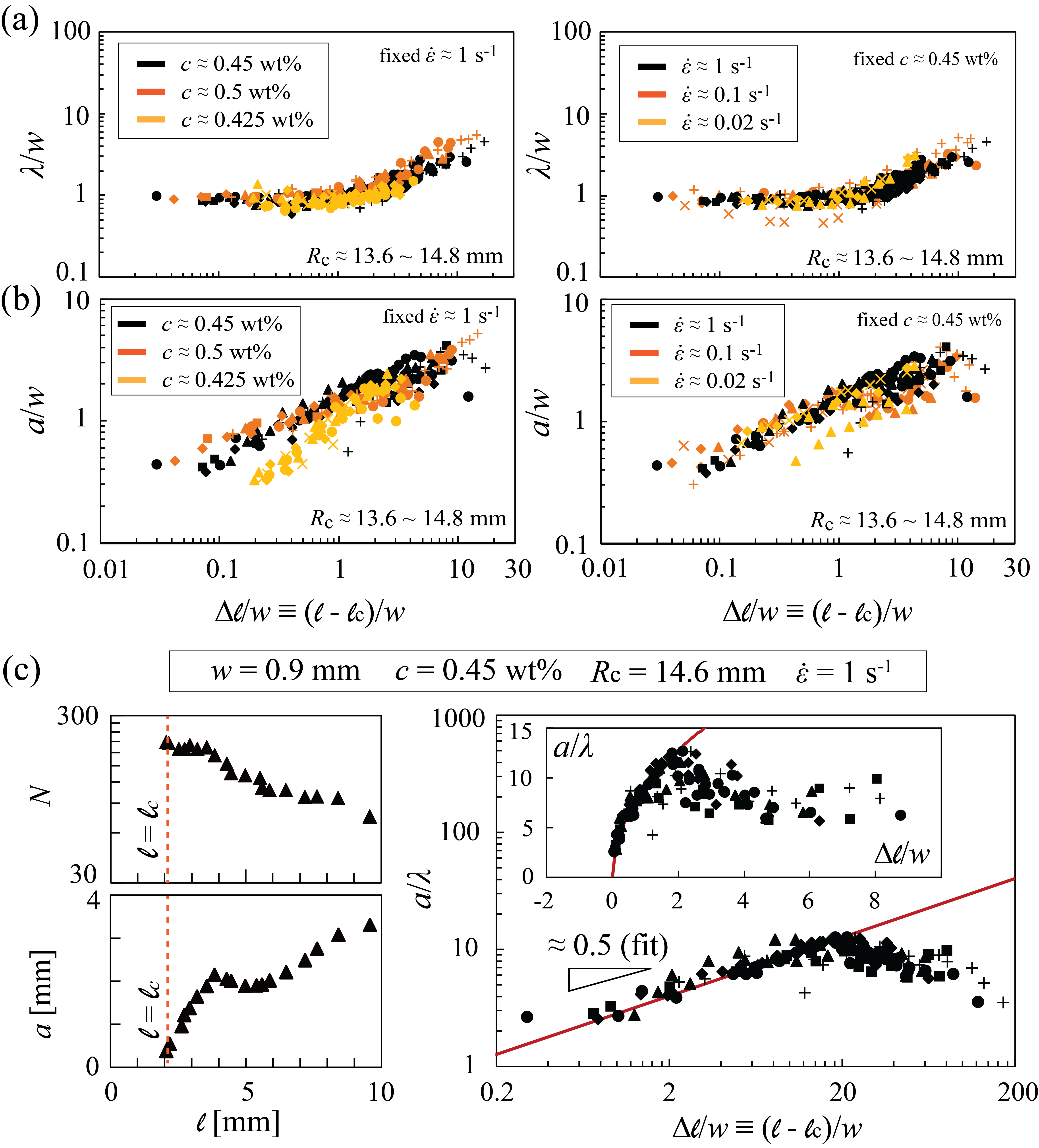}% Here is how to import EPS art
\caption{\label{fig:FigureS7}Post-instability characteristics in $\lambda$ and $a$. (a, b) Supplements to the data shown in Fig. 4(a) in the main text. Left two plots measure $\lambda/w$ and $a/w$ as function of the excess strain $\Delta\ell/w=(\ell-\ell_c)/w$ for fixed extrusion rate $\dot\varepsilon$. The right two plots measure the same thing with a fixed crosslinker concentration $c=0.45$ wt\%. (c) Measured post-instability trends in $N$ and $a$ of $c=0.45$ wt\% gel extruded at $\dot\varepsilon=1$ s$^{-1}$ and in $w=0.9$ mm, $R_o=15.5$ mm annulus. The aspect ratio of the furrows $a/\lambda$ grows nonlinearly as the excess strain $\Delta\ell/w$ increases. Data in this plot were taken with annulus with varied $w$ but constant $R_o$. (inset) $a/\lambda$ versus $\Delta\ell/w$ in linear-linear plot. Line in red is fitted power-law curve with an exponent of $0.5$.}
\end{figure*}

In Fig. \ref{fig:FigureS7}, we present additional set of data in conjunction to Fig. 4(a) shown in the main text. We measured the wavelength $\lambda$ and the furrow depth $a$ for different values of the strain rate $\dot\varepsilon=[0.02, 0.1, 1] \text{ s}^{-1}$ and crosslinker concentration $c=[0.425, 0.45, 0.5]$ wt\%, as a function of the excess strain defined as $\Delta\ell/w=(\ell-\ell_c)/w$. Regardless of $\dot\varepsilon$ and $c$, the furrows evolve longer wavelengths once $(\ell-\ell_c)>w$. However, we observed no strong dependence of $a(\ell(t))$ on $\dot\varepsilon$, in contrast to the result shown in Fig. 4(a), which supports our conclusion that the instability is elasticity driven. Note that we measure $a$ from the time when the apex of the curved front furrows downward. 

To highlight the evolution of the individual furrows, we plot the aspect ratio of protuberances $a/\lambda$ as a function of the excess strain in Fig. \ref{fig:FigureS7}(c). In this representative case, we have $c=0.45$ wt\% gel extruded at $\dot\varepsilon=1$ s$^{-1}$ inside the $w=0.9$ mm, $R_o=15.5$ mm annulus. The aspect ratio of the furrows $a/\lambda$ grows nonlinearly with $\Delta\ell/w$ in a power-law fashion ($a/\lambda\sim(\Delta\ell/w)^\alpha$, where $\alpha$ is some exponent). For $c=0.425, 0.45, 0.5$ wt\%, the fitted exponent were $\alpha=0.79, 0.5, 0.25$, respectively. \\

\section{Experimental methods}
\textbf{\textit{Syringe fabrication – }}We fabricated cylindrical annulus at varying gaps using off-the-shelf cast acrylics. Extruded acrylics were not suitable for our experiment because the extruded materials typically have rough extrusion lines on the surface of the material. We lathed the cast acrylic rod (for example, McMaster Carr item No. 8528K53) down to varying diameters to fabricate the inner cylinder in the annulus. These machined parts were polished with 1000 grade sand paper with water, and ultrasoft jewelry polishing cloth. The outer, hollow cylinder part of the annulus was fabricated from McMaster Carr, for example, item No. 8486K331. We 3D-printed the syringe plunger (Form 2 printer, FormLabs) and used an O-ring of appropriate size to seal the gap between the 3D printed parts and the machined hollow cylinders. \\

\begin{figure*}
\includegraphics{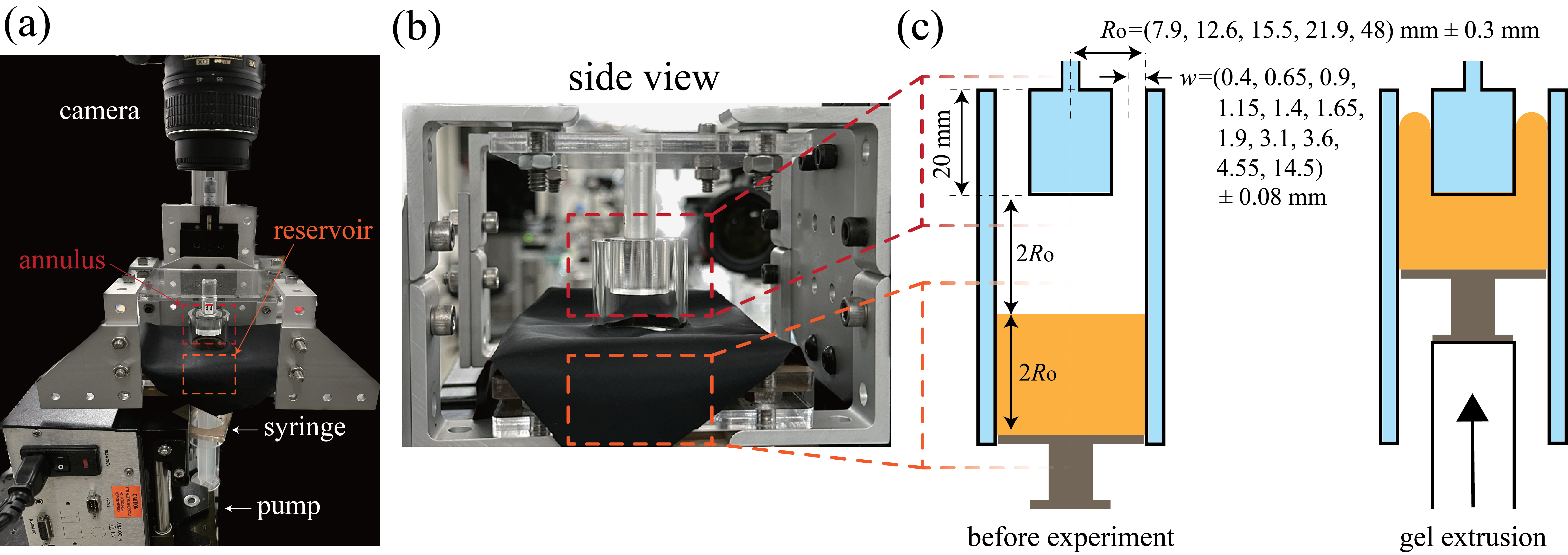}% Here is how to import EPS art
\caption{\label{fig:FigureS8}Experimental setup. (a) Overall view of the setup. (b) Side view of the experiment box. The gel reservoir (region inside the dotted orange box) is covered underneath the black cloth. (c) The cast gel has height/diameter ratio of close to 1. Also, the free surface of the gel in the cast state is positioned approximately $2R_o$ below from the annular section.}
\end{figure*}

\textbf{\textit{Before the extrusion – }}We cast the gel in the gel reservoir (see Fig. \ref{fig:FigureS8}) and let it cure for overnight (approximately 15 hours). The height of the cast gel is kept approximately the same as the diameter ($2R_o$) of the channel. Also, the distance between the start of the annular section to the free surface of the cast gel was about $1.5\sim2R_o$. We kept this ratio throughout the experiments for consistency, but changing this initial distance may provide new insights. This distance must not be larger than $2R_o$ as the gel can develop furrowing instability just by flowing inside a hollow cylinder ($\ell_c\approx 2.7 R_o$; see Fig. 2(a) in the main text). \\

\textbf{\textit{During the extrusion – }}Once the gel is fully cured, the gel was pushed vertically upward with a syringe pump (PHD Ultra, Harvard Apparatus) at varying flow rates as input to the pump. Given a field of view of our lens, the side view video filmed the region of an arclength of 12 mm to 16 mm. Because the feature size of our interest was typically \~ 1 mm or less in its size, we could take average of the 10 to 20 furrows in each experiment. For the side view, we used Nikon D3300 with a lens with 200mm focal length. For the top view, we used Nikon D5100 with an 18-55mm lens. After each experiment, the acrylic parts were disassembled and were thoroughly cleaned with ethanol to remove PDMS residue.

The furrowing instability begins from the inner boundary ($R_i$) and propagates outward to ($R_o$). We defined $\ell_c$ as the position of the apex of the free surface when the apex starts to furrow downward, instead of when the inner contact line starts to furrow. This is because the camera is positioned outside (looking into the annulus from the side) and the view is distorted if one would try to observe the inner contact line through the curved surface. Also, as the propagation of the instability is much faster than the time scale of the extrusion (see Fig. 3(c) in the main text), this brief lag of time for the instability to propagate from $R_i$ to the apex did not affect our result. Likewise, the furrow depth $a$ was measured from when the apex starts to furrow downward.\\

\section{Supporting videos}
Supporting video 1: a side view of the $c=0.45$ wt\% gel extruded at $\dot\varepsilon=1$ s$^{-1}$ in an $w=0.65$ mm and $R_o=15.5$ mm shown in the Fig. 1(b) in the main text. The video was filmed at real-time.   \\

Supporting video 2: a side view of the $c=0.45$ wt\% gel extruded at $\dot\varepsilon=0.1$ s$^{-1}$ in an $w=1.65$ mm and $R_o=15.5$ mm shown in the Fig. 3(a) in the main text. The video was filmed at real-time. This video shows flow of the density-matched Polystyrene beads with the gel. \\

Supporting video 3: a top view of the $c=0.50$ wt\% gel extruded at $\dot\varepsilon=0.1$ s$^{-1}$ in an $w=4.55$ mm and $R_o=15.5$ mm shown in the Fig. 3(c) in the main text. The video was filmed at real-time. This video shows radial propagation of the instability from the inner boundary to the outer boundary, and also shows partial relaxation of the buckled surface after the extrusion had stopped.\\

%\bibliography{References.bib}% Produces the bibliography via BibTeX.

\newpage

